\documentclass[envcountsame]{article}
\usepackage{times,amsmath,amssymb}
\usepackage{tikz}
\newcommand{\V}{
\draw (-3,4) -- (0,0) -- (3,4);
\draw (-4,4) -- (-2,4);
\draw (4,4) -- (2,4);
}

\newcommand{\wt}{
\draw (-3,4) -- (0,0) -- (3,4) -- (-3,4);
}

\newcommand{\bt}{
\draw[fill=black] (-3,4) -- (0,0) -- (3,4) -- (-3,4);
}

\begin{document}

\title{The Trees of Hanoi}
\author{
Joost Engelfriet \\ \\
\small Leiden Institute of Advanced Computer Science \\
\small Leiden University, The Netherlands \\
\small \texttt{j.engelfriet@liacs.leidenuniv.nl}
}

\date{{}}

\maketitle


\begin{abstract}
The game of the Towers of Hanoi is generalized to binary trees. 
First, a straightforward solution of the game is discussed. 
Second, a shorter solution is presented, which is then shown to be optimal. 
\end{abstract}

\vspace{2cm}
{\bf Contents} \hspace{5.9cm} page

\bigskip
The game, and its solution \hspace{3.8cm} 2

\medskip
Acknowledgement \hspace{4.855cm} 9

\medskip
References \hspace{5.75cm} 10

\medskip
Epilogue \hspace{6.053cm} 11

\medskip
References \hspace{5.75cm} 11

\vspace{2cm}
This note is a slightly revised version 
of a note from 1981, see the Epilogue.

\newpage
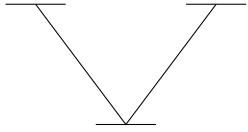
\begin{center}
\begin{tikzpicture}
\begin{scope}[scale=0.8]
\draw (0,2) -- (1.5,0) -- (3,2);
\draw (1,0) -- (2,0);
\draw (-0.5,2) -- (0.5,2);
\draw (2.5,2) -- (3.5,2);
\end{scope}
\end{tikzpicture}
\end{center}

\section*{The game, and its solution}

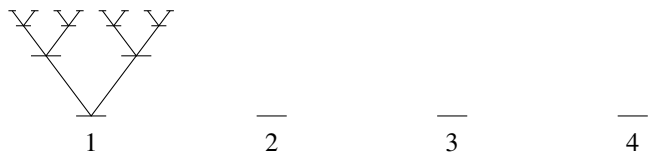
\begin{figure}[b]
\begin{tikzpicture}[scale = 0.2]
\node at (0,10) { };
\draw (-1,0) -- (1,0);
\node [below] at (0,-0.5) {1};
\draw (11,0) -- (13,0);
\node [below] at (12,-0.5) {2};
\draw (23,0) -- (25,0);
\node [below] at (24,-0.5) {3};
\draw (35,0) -- (37,0);
\node [below] at (36,-0.5) {4};

\V
\begin{scope}[shift={(-3,4)}, scale=0.5]
\V
\begin{scope}[shift={(-3,4)}, scale=0.5]
\V
\end{scope}
\begin{scope}[shift={(3,4)}, scale=0.5]
\V
\end{scope}
\end{scope}
\begin{scope}[shift={(3,4)}, scale=0.5]
\V
\begin{scope}[shift={(-3,4)}, scale=0.5]
\V
\end{scope}
\begin{scope}[shift={(3,4)}, scale=0.5]
\V
\end{scope}
\end{scope}
\end{tikzpicture}
  \caption{Initial configuration with full binary tree of height 3.}
  \label{fig:init}
\end{figure}

The well-known game of the Towers of Hanoi can be generalized in such a way 
that a full binary tree is moved from one place to another, 
rather than a pile of discs. Moreover, there are four places in total, 
rather than three in the case of discs (where places are called pegs). 
The initial configuration with a full binary tree of height 3 is shown in Fig.~\ref{fig:init}. 
We assume that the sons of a node of the full binary tree 
are of the same size but smaller than their father. More precisely, if the tree is of height $n$, 
then there are $2^{i-1}$ nodes of the same size at the $i$th level, $1\leq i\leq n$, 
and the size decreases from level to level. The nodes at each level are indistinguishable.

At each step of the game one may move a leaf, i.e., a node without sons, 
from one position to another (free) position where it is again a leaf,
always under the condition that it should be smaller than its father (if it has one). 
In particular a node may be moved to an unoccupied place, or it may be moved 
to be the other son of its father (if that position is free). 
Thus, at each moment of time, each nonempty place contains a (not necessary full) binary tree
in which each son is of smaller size than its father (but not necessarily of the 
next smaller size). The two sons of a father need not be of the same size. 

To have in mind a concrete picture of the game one may visualize each node as a 
\begin{tikzpicture}[scale = 0.07]
\draw (-1,0) -- (1,0);
\V
\end{tikzpicture}
\!-shaped object: on each of its two legs another (smaller) such object may be placed.
Numbering the four places from 1 to 4, each possible position of a node during the game 
can be coded as a string $kD_1D_2\cdots D_m$ 
with $1\leq k\leq 4$, $0\leq m\leq n-1$, and $D_i\in\{L,R\}$ 
where $L$ stands for `left' and $R$ for `right'. 
Thus, in Fig.~\ref{fig:confs}(a) there are nodes at positions 
$1$, $1R$, $1RL$, $2$, $2L$, $2R$, and $4$. 
By `move($x,y$)', where $x$~and~$y$ are possible positions, we denote a move of the node 
at position $x$ to position~$y$. As an example, in Fig.~\ref{fig:confs}(a), after move($1RL,1L$)
we can do move($1R,3$), and the configuration will be as in Fig.~\ref{fig:confs}(b). 
\begin{figure}[t]
\begin{tikzpicture}[scale = 0.2]
\draw (-1,0) -- (1,0);
\node [below] at (0,-0.5) {1};
\draw (11,0) -- (13,0);
\node [below] at (12,-0.5) {2};
\draw (23,0) -- (25,0);
\node [below] at (24,-0.5) {3};
\draw (35,0) -- (37,0);
\node [below] at (36,-0.5) {4};

\V
\begin{scope}[shift={(3,4)}, scale=0.5]
\V
\begin{scope}[shift={(-3,4)}, scale=0.5]
\V
\end{scope}
\end{scope}

\begin{scope}[shift={(12,0)}, scale=0.5]

\V
\begin{scope}[shift={(-3,4)}, scale=0.5]
\V
\end{scope}
\begin{scope}[shift={(3,4)}, scale=0.5]
\V
\end{scope}

\end{scope}

\begin{scope}[shift={(36,0)},scale=0.25]
\V
\end{scope}
\end{tikzpicture}

\vspace{0.5cm}
\hspace{4cm} (a)
\vspace{1cm}

\begin{tikzpicture}[scale = 0.2]
\draw (-1,0) -- (1,0);
\node [below] at (0,-0.5) {1};
\draw (11,0) -- (13,0);
\node [below] at (12,-0.5) {2};
\draw (23,0) -- (25,0);
\node [below] at (24,-0.5) {3};
\draw (35,0) -- (37,0);
\node [below] at (36,-0.5) {4};

\V
\begin{scope}[shift={(-3,4)}, scale=0.25]
\V
\end{scope}

\begin{scope}[shift={(12,0)},scale=0.5]

\V
\begin{scope}[shift={(-3,4)}, scale=0.5]
\V
\end{scope}
\begin{scope}[shift={(3,4)}, scale=0.5]
\V
\end{scope}

\end{scope}

\begin{scope}[shift={(24,0)}, scale=0.5]
\V
\end{scope}

\begin{scope}[shift={(36,0)}, scale=0.25]
\V
\end{scope}

\end{tikzpicture}

\vspace{0.5cm}
\hspace{4cm} (b)
  \caption{Configurations.}
  \label{fig:confs}
\end{figure}

Solving the problem for the slightly more general case where the places are actually positions
in a game with a larger $n$, it is easy to see (as in the case of the Towers of Hanoi) 
that the following recursive Pascal-like procedure $t$ does the job 
(where the type `position' is as explained above).
A call $t(n,a,b,c,d)$ will move the full binary tree of height $n$ from $a$ to $b$, via $c$ and $d$. 

\bigskip
\noindent
{\bf procedure} $t(n: \text{integer}; \;a,b,c,d: \text{position})$;\\

\noindent
{\bf begin} \;\;{\bf if} $n>0$ \;{\bf then} 
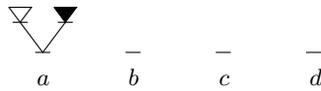
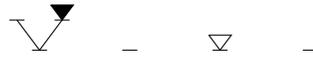
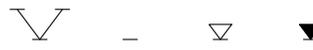
\hspace{1.45cm}
\begin{tikzpicture}[scale = 0.1]
\draw (-1,0) -- (1,0);
\draw (11,0) -- (13,0);
\draw (23,0) -- (25,0);
\draw (35,0) -- (37,0);
\V
\begin{scope}[shift={(-3,4)}, scale=0.5]
\wt
\end{scope}
\begin{scope}[shift={(3,4)}, scale=0.5]
\bt
\end{scope}
\end{tikzpicture}

\hspace{4.5cm} {\small $a$}  
\hspace{0.87cm} {\small $b$} 
\hspace{0.89cm} {\small $c$} 
\hspace{0.89cm} {\small $d$}

{\bf begin}

\hspace{0.3cm}$t(n-1,aL,c,b,d)$;
\hspace{1cm}
\begin{tikzpicture}[scale = 0.1]
\draw (-1,0) -- (1,0);
\draw (11,0) -- (13,0);
\draw (23,0) -- (25,0);
\draw (35,0) -- (37,0);
\V
\begin{scope}[shift={(24,0)}, scale=0.5]
\wt
\end{scope}
\begin{scope}[shift={(3,4)}, scale=0.5]
\bt
\end{scope}
\end{tikzpicture}\\

\hspace{0.3cm}$t(n-1,aR,d,aL,b)$;
\hspace{0.7cm}
\begin{tikzpicture}[scale = 0.1]
\draw (-1,0) -- (1,0);
\draw (11,0) -- (13,0);
\draw (23,0) -- (25,0);
\draw (35,0) -- (37,0);
\V
\begin{scope}[shift={(24,0)}, scale=0.5]
\wt
\end{scope}
\begin{scope}[shift={(36,0)}, scale=0.5]
\bt
\end{scope}
\end{tikzpicture}\\

\hspace{0.3cm}move$(a,b)$;
\hspace{2.5cm}
\begin{tikzpicture}[scale = 0.1]
\draw (-1,0) -- (1,0);
\draw (11,0) -- (13,0);
\draw (23,0) -- (25,0);
\draw (35,0) -- (37,0);
\begin{scope}[shift={(12,0)}]
\V
\end{scope}
\begin{scope}[shift={(24,0)}, scale=0.5]
\wt
\end{scope}
\begin{scope}[shift={(36,0)}, scale=0.5]
\bt
\end{scope}
\end{tikzpicture}\\

\hspace{0.3cm}$t(n-1,c,bL,a,bR)$;
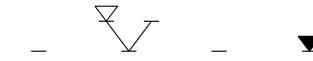
\hspace{1.05cm}
\begin{tikzpicture}[scale = 0.1]
\draw (-1,0) -- (1,0);
\draw (11,0) -- (13,0);
\draw (23,0) -- (25,0);
\draw (35,0) -- (37,0);
\begin{scope}[shift={(12,0)}]
\V
\end{scope}
\begin{scope}[shift={(9,4)}, scale=0.5]
\wt
\end{scope}
\begin{scope}[shift={(36,0)}, scale=0.5]
\bt
\end{scope}
\end{tikzpicture}\\

\hspace{0.3cm}$t(n-1,d,bR,a,c)$
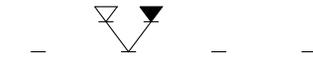
\hspace{1.35cm}
\begin{tikzpicture}[scale = 0.1]
\draw (-1,0) -- (1,0);
\draw (11,0) -- (13,0);
\draw (23,0) -- (25,0);
\draw (35,0) -- (37,0);
\begin{scope}[shift={(12,0)}]
\V
\end{scope}
\begin{scope}[shift={(9,4)}, scale=0.5]
\wt
\end{scope}
\begin{scope}[shift={(15,4)}, scale=0.5]
\bt
\end{scope}
\end{tikzpicture}\\

{\bf end}\\

\noindent
{\bf end.}

\bigskip
\noindent
At the right are shown the configurations after each statement, 
where \,\begin{tikzpicture}[scale=0.05] \wt \end{tikzpicture}\, indicates the left subtree 
and \,\begin{tikzpicture}[scale=0.05] \bt \end{tikzpicture}\, the right subtree of height $n-1$. 

Let $t_n$ denote the number of moves executed by a call $t(n,\dots)$. 
Clearly $t_0=0$, $t_1=1$, and $t_n=4t_{n-1}+1$. 
Hence $t_n= 4^{n-1}+\cdots+4^2+4+1=\frac{1}{3}(4^n-1)$. 
Note that the number of nodes in the full binary tree of height $n$ is $2^n-1$. 
Thus the number of moves is quadratic in the number of nodes. 

It is easy to show that the moves executed by the call $t(n,1,2,3,4)$ satisfy the following additional 
condition: a node is always put on top of one of its original ancestors (or directly on a place). 
More formally, if node $v_1$ is put on one of the legs of node $v_2$, 
then $p(v_2)$ is a proper prefix of $p(v_1)$, where $p(v)$ denotes the position of node $v$ 
in the initial configuration. Adding this condition as a requirement to the game it is easy to prove
that $t_n$ is the minimal number of moves needed. This can be seen by the usual bottleneck argument 
(see~\cite{woo80}), as follows. For a given sequence $w$ of moves from the initial to the final configuration
(for a full binary tree of height $n$), consider the first move of the largest node; 
\vspace{0.1cm} thus there is a prefix $u$ of $w$ leading from 
\begin{tikzpicture}[scale = 0.1]
\draw (-1,0) -- (1,0);
\draw (11,0) -- (13,0);
\draw (23,0) -- (25,0);
\draw (35,0) -- (37,0);
\V
\begin{scope}[shift={(-3,4)}, scale=0.5]
\wt
\end{scope}
\begin{scope}[shift={(3,4)}, scale=0.5]
\bt
\end{scope}
\end{tikzpicture}
$\quad$ to $\quad$
\begin{tikzpicture}[scale = 0.1]
\draw (-1,0) -- (1,0);
\draw (11,0) -- (13,0);
\draw (23,0) -- (25,0);
\draw (35,0) -- (37,0);
\V
\begin{scope}[shift={(24,0)}, scale=0.5]
\wt
\end{scope}
\begin{scope}[shift={(36,0)}, scale=0.5]
\bt
\end{scope}
\end{tikzpicture}
$\quad$ (or \vspace{0.1cm} any other configuration obtained from this one 
by permuting places $2$, $3$, and $4$), \vspace{0.1cm}after which 
\begin{tikzpicture}[scale = 0.1]
\draw (-1,0) -- (1,0);
\V
\end{tikzpicture}
moves for the first time. 
In this prefix $u$ of $w$ only moves with ``white'' nodes and ``black'' nodes are made
(from the left and right subtree, respectively). Moreover, due to the additional requirement,
no black node is ever on a white node, or vice versa. This means that by restricting attention 
to all white moves (disregarding the black nodes) one obtains a sequence of moves for 
moving a full binary tree of height $n-1$. Since the same is true for the black moves, 
it follows that (reasoning by induction) $u$ contains at least $2t_{n-1}$ moves. 
Similarly, considering the last move of the largest node, 
a postfix of $w$ of length at least $2t_{n-1}$ is obtained (disjoint with $u$). 
And so the length of $w$ is at least $2t_{n-1}+1+2t_{n-1}=4t_{n-1}+1=t_n$. 

We will now show that, without the above additional condition, we can do with less moves. 
For $n=3$, $t$ needs 21 moves ($t_3=21$), but actually 19 moves suffice 
as can be seen from the following sequence of configurations: 

\bigskip
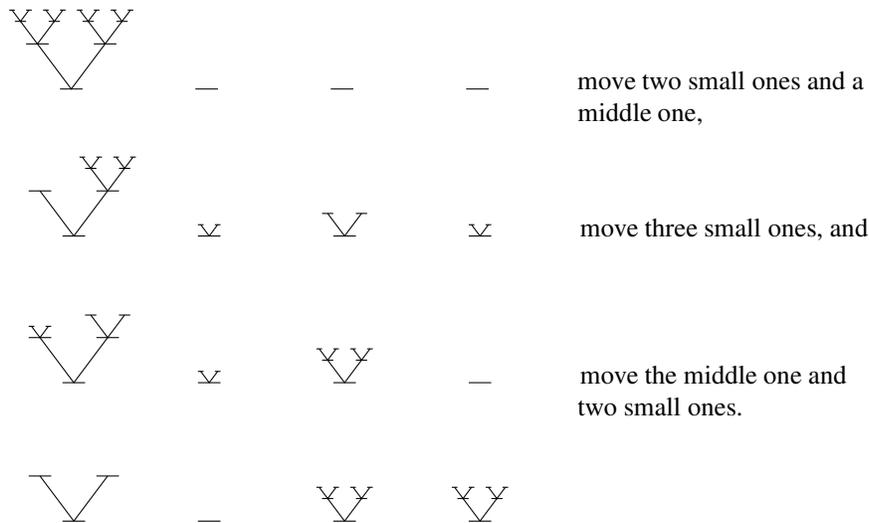
\begin{tikzpicture}[scale = 0.15]
\draw (-1,0) -- (1,0);
\draw (11,0) -- (13,0);
\draw (23,0) -- (25,0);
\draw (35,0) -- (37,0);

\V
\begin{scope}[shift={(-3,4)}, scale=0.5]
\V
\begin{scope}[shift={(-3,4)}, scale=0.5]
\V
\end{scope}
\begin{scope}[shift={(3,4)}, scale=0.5]
\V
\end{scope}
\end{scope}
\begin{scope}[shift={(3,4)}, scale=0.5]
\V
\begin{scope}[shift={(-3,4)}, scale=0.5]
\V
\end{scope}
\begin{scope}[shift={(3,4)}, scale=0.5]
\V
\end{scope}
\end{scope}
\end{tikzpicture}
$\hspace{1cm}$
move two small ones and a 

$\hspace{7.48cm}$
middle one,

\bigskip
\hspace{0.17cm}
\begin{tikzpicture}[scale = 0.15]
\draw (-1,0) -- (1,0);
\draw (11,0) -- (13,0);
\draw (23,0) -- (25,0);
\draw (35,0) -- (37,0);

\V
\begin{scope}[shift={(3,4)}, scale=0.5]
\V
\begin{scope}[shift={(-3,4)}, scale=0.5]
\V
\end{scope}
\begin{scope}[shift={(3,4)}, scale=0.5]
\V
\end{scope}
\end{scope}

\begin{scope}[shift={(12,0)}, scale=0.25]
\V
\end{scope}

\begin{scope}[shift={(24,0)}, scale=0.5]
\V
\end{scope}

\begin{scope}[shift={(36,0)}, scale=0.25]
\V
\end{scope}
\end{tikzpicture}
$\hspace{1cm}$
move three small ones, and

\bigskip
\bigskip
\medskip
\hspace{0.17cm}
\begin{tikzpicture}[scale = 0.15]
\draw (-1,0) -- (1,0);
\draw (11,0) -- (13,0);
\draw (23,0) -- (25,0);
\draw (35,0) -- (37,0);

\V
\begin{scope}[shift={(-3,4)}, scale=0.25]
\V
\end{scope}
\begin{scope}[shift={(3,4)}, scale=0.5]
\V
\end{scope}

\begin{scope}[shift={(12,0)}, scale=0.25]
\V
\end{scope}

\begin{scope}[shift={(24,0)}, scale=0.5]
\V
\begin{scope}[shift={(-3,4)}, scale=0.5]
\V
\end{scope}
\begin{scope}[shift={(3,4)}, scale=0.5]
\V
\end{scope}
\end{scope}
\end{tikzpicture}
$\hspace{1cm}$
move the middle one and 

$\hspace{7.48cm}$
two small ones.

\bigskip
\bigskip
\hspace{0.17cm}
\begin{tikzpicture}[scale = 0.15]
\draw (-1,0) -- (1,0);
\draw (11,0) -- (13,0);
\draw (23,0) -- (25,0);
\draw (35,0) -- (37,0);

\V

\begin{scope}[shift={(24,0)}, scale=0.5]
\V
\begin{scope}[shift={(-3,4)}, scale=0.5]
\V
\end{scope}
\begin{scope}[shift={(3,4)}, scale=0.5]
\V
\end{scope}
\end{scope}

\begin{scope}[shift={(36,0)}, scale=0.5]
\V
\begin{scope}[shift={(-3,4)}, scale=0.5]
\V
\end{scope}
\begin{scope}[shift={(3,4)}, scale=0.5]
\V
\end{scope}
\end{scope}
\end{tikzpicture}

\noindent
This takes 9 moves. Now the large one can be moved and 
a symmetric sequence of moves can be used to build up the tree. 
Together we have uses $9+1+9=19$ moves. 

This idea is used in the following Pascal-like recursive procedures $f$, $g$, and $h$, 
where $f(n,a,b,c,d)$
moves the tree of height $n$ from $a$ to $b$, via $c$ and $d$ (i.e., $f$~solves our problem), 
$g(n,a,b,c,d,e)$ moves \emph{two} trees of height $n$ from positions $a$ and $b$
to positions $c$ and $d$, via $e$, and finally, $h(n,a,b,c,x,y,z)$ 
moves \emph{three} trees of height~$n$ from positions $a,b,c$ to positions $x,y,z$. 
In the pictures, \,\begin{tikzpicture}[scale=0.05] \wt \end{tikzpicture}\,
denotes a subtree of height~$n-1$. 

\bigskip
\noindent
{\bf procedure} $f(n: \text{integer}; \;a,b,c,d: \text{position})$;\\

\noindent
{\bf begin} \;\;{\bf if} $n>0$ \;{\bf then} 
\hspace{1.85cm}
\begin{tikzpicture}[scale = 0.1]
\draw (-1,0) -- (1,0);
\draw (11,0) -- (13,0);
\draw (23,0) -- (25,0);
\draw (35,0) -- (37,0);
\V
\begin{scope}[shift={(-3,4)}, scale=0.5]
\wt
\end{scope}
\begin{scope}[shift={(3,4)}, scale=0.5]
\wt
\end{scope}
\end{tikzpicture}

\hspace{4.9cm} {\small $a$}  
\hspace{0.87cm} {\small $b$} 
\hspace{0.89cm} {\small $c$} 
\hspace{0.89cm} {\small $d$}

{\bf begin}

\hspace{0.3cm}$g(n-1,aL,aR,c,d,b)$;
\hspace{0.7cm}
\begin{tikzpicture}[scale = 0.1]
\draw (-1,0) -- (1,0);
\draw (11,0) -- (13,0);
\draw (23,0) -- (25,0);
\draw (35,0) -- (37,0);
\V
\begin{scope}[shift={(24,0)}, scale=0.5]
\wt
\end{scope}
\begin{scope}[shift={(36,0)}, scale=0.5]
\wt
\end{scope}
\end{tikzpicture}\\

\hspace{0.3cm}move$(a,b)$;
\hspace{2.9cm}
\begin{tikzpicture}[scale = 0.1]
\draw (-1,0) -- (1,0);
\draw (11,0) -- (13,0);
\draw (23,0) -- (25,0);
\draw (35,0) -- (37,0);
\begin{scope}[shift={(12,0)}]
\V
\end{scope}
\begin{scope}[shift={(24,0)}, scale=0.5]
\wt
\end{scope}
\begin{scope}[shift={(36,0)}, scale=0.5]
\wt
\end{scope}
\end{tikzpicture}\\

\hspace{0.3cm}$g(n-1,c,d,bL,bR,a)$
\hspace{1.15cm}
\begin{tikzpicture}[scale = 0.1]
\draw (-1,0) -- (1,0);
\draw (11,0) -- (13,0);
\draw (23,0) -- (25,0);
\draw (35,0) -- (37,0);
\begin{scope}[shift={(12,0)}]
\V
\end{scope}
\begin{scope}[shift={(9,4)}, scale=0.5]
\wt
\end{scope}
\begin{scope}[shift={(15,4)}, scale=0.5]
\wt
\end{scope}
\end{tikzpicture}\\

{\bf end}

\smallskip
\noindent
{\bf end;}

\vspace{0.8cm}
\noindent
{\bf procedure} $g(n: \text{integer}; \;a,b,c,d,e: \text{position})$;\\

\noindent
{\bf begin} \;\;{\bf if} $n>0$ \;{\bf then} 
\hspace{2.35cm}
\begin{tikzpicture}[scale = 0.1]
\draw (-1,0) -- (1,0);
\draw (11,0) -- (13,0);
\draw (23,0) -- (25,0);
\draw (35,0) -- (37,0);
\draw (47,0) -- (49,0);
\V
\begin{scope}[shift={(-3,4)}, scale=0.5]
\wt
\end{scope}
\begin{scope}[shift={(3,4)}, scale=0.5]
\wt
\end{scope}
\begin{scope}[shift={(12,0)}]
\V
\begin{scope}[shift={(-3,4)}, scale=0.5]
\wt
\end{scope}
\begin{scope}[shift={(3,4)}, scale=0.5]
\wt
\end{scope}
\end{scope}
\end{tikzpicture}

\hspace{5.4cm} {\small $a$}  
\hspace{0.87cm} {\small $b$} 
\hspace{0.89cm} {\small $c$} 
\hspace{0.89cm} {\small $d$}
\hspace{0.89cm} {\small $e$}

{\bf begin}

\hspace{0.3cm}$g(n-1,aL,aR,d,e,c)$;
\hspace{1.2cm}
\begin{tikzpicture}[scale = 0.1]
\draw (-1,0) -- (1,0);
\draw (11,0) -- (13,0);
\draw (23,0) -- (25,0);
\draw (35,0) -- (37,0);
\draw (47,0) -- (49,0);
\V
\begin{scope}[shift={(12,0)}]
\V
\begin{scope}[shift={(-3,4)}, scale=0.5]
\wt
\end{scope}
\begin{scope}[shift={(3,4)}, scale=0.5]
\wt
\end{scope}
\end{scope}

\begin{scope}[shift={(36,0)}, scale=0.5]
\wt
\end{scope}
\begin{scope}[shift={(48,0)}, scale=0.5]
\wt
\end{scope}
\end{tikzpicture}\\

\hspace{0.3cm}move$(a,c)$;
\hspace{3.4cm}
\begin{tikzpicture}[scale = 0.1]
\draw (-1,0) -- (1,0);
\draw (11,0) -- (13,0);
\draw (23,0) -- (25,0);
\draw (35,0) -- (37,0);
\draw (47,0) -- (49,0);

\begin{scope}[shift={(12,0)}]
\V
\begin{scope}[shift={(-3,4)}, scale=0.5]
\wt
\end{scope}
\begin{scope}[shift={(3,4)}, scale=0.5]
\wt
\end{scope}
\end{scope}

\begin{scope}[shift={(24,0)}]
\V
\end{scope}

\begin{scope}[shift={(36,0)}, scale=0.5]
\wt
\end{scope}
\begin{scope}[shift={(48,0)}, scale=0.5]
\wt
\end{scope}
\end{tikzpicture}\\

\hspace{0.3cm}$h(n-1,bL,bR,d,a,cL,cR)$;
\hspace{0.7cm}
\begin{tikzpicture}[scale = 0.1]
\draw (-1,0) -- (1,0);
\draw (11,0) -- (13,0);
\draw (23,0) -- (25,0);
\draw (35,0) -- (37,0);
\draw (47,0) -- (49,0);

\begin{scope}[scale=0.5]
\wt
\end{scope}

\begin{scope}[shift={(12,0)}]
\V
\end{scope}

\begin{scope}[shift={(24,0)}]
\V
\begin{scope}[shift={(-3,4)}, scale=0.5]
\wt
\end{scope}
\begin{scope}[shift={(3,4)}, scale=0.5]
\wt
\end{scope}
\end{scope}

\begin{scope}[shift={(48,0)}, scale=0.5]
\wt
\end{scope}
\end{tikzpicture}\\

\hspace{0.3cm}move$(b,d)$;
\hspace{3.4cm}
\begin{tikzpicture}[scale = 0.1]
\draw (-1,0) -- (1,0);
\draw (11,0) -- (13,0);
\draw (23,0) -- (25,0);
\draw (35,0) -- (37,0);
\draw (47,0) -- (49,0);

\begin{scope}[scale=0.5]
\wt
\end{scope}

\begin{scope}[shift={(24,0)}]
\V
\begin{scope}[shift={(-3,4)}, scale=0.5]
\wt
\end{scope}
\begin{scope}[shift={(3,4)}, scale=0.5]
\wt
\end{scope}
\end{scope}

\begin{scope}[shift={(36,0)}]
\V
\end{scope}

\begin{scope}[shift={(48,0)}, scale=0.5]
\wt
\end{scope}
\end{tikzpicture}\\

\hspace{0.3cm}$g(n-1,a,e,dL,dR,b)$;
\hspace{1.55cm}
\begin{tikzpicture}[scale = 0.1]
\draw (-1,0) -- (1,0);
\draw (11,0) -- (13,0);
\draw (23,0) -- (25,0);
\draw (35,0) -- (37,0);
\draw (47,0) -- (49,0);

\begin{scope}[shift={(24,0)}]
\V
\begin{scope}[shift={(-3,4)}, scale=0.5]
\wt
\end{scope}
\begin{scope}[shift={(3,4)}, scale=0.5]
\wt
\end{scope}
\end{scope}

\begin{scope}[shift={(36,0)}]
\V
\begin{scope}[shift={(-3,4)}, scale=0.5]
\wt
\end{scope}
\begin{scope}[shift={(3,4)}, scale=0.5]
\wt
\end{scope}
\end{scope}
\end{tikzpicture}\\

{\bf end}

\smallskip
\noindent
{\bf end;}

\noindent
{\bf procedure} $h(n: \text{integer}; \;a,b,c,x,y,z: \text{position})$;\\

\noindent
{\bf begin} \;\;{\bf if} $n>0$ \;{\bf then} 
\hspace{2.39cm}
\begin{tikzpicture}[scale = 0.1]
\draw (-1,0) -- (1,0);
\draw (11,0) -- (13,0);
\draw (23,0) -- (25,0);
\draw (35,0) -- (37,0);
\draw (47,0) -- (49,0);
\draw (59,0) -- (61,0);
\V
\begin{scope}[shift={(-3,4)}, scale=0.5]
\wt
\end{scope}
\begin{scope}[shift={(3,4)}, scale=0.5]
\wt
\end{scope}
\begin{scope}[shift={(12,0)}]
\V
\begin{scope}[shift={(-3,4)}, scale=0.5]
\wt
\end{scope}
\begin{scope}[shift={(3,4)}, scale=0.5]
\wt
\end{scope}
\end{scope}
\begin{scope}[shift={(24,0)}]
\V
\begin{scope}[shift={(-3,4)}, scale=0.5]
\wt
\end{scope}
\begin{scope}[shift={(3,4)}, scale=0.5]
\wt
\end{scope}
\end{scope}
\end{tikzpicture}

\hspace{5.4cm} {\small $a$}  
\hspace{0.87cm} {\small $b$} 
\hspace{0.89cm} {\small $c$} 
\hspace{0.89cm} {\small $x$}
\hspace{0.89cm} {\small $y$}
\hspace{0.89cm} {\small $z$}

{\bf begin}

\hspace{0.3cm}$g(n-1,aL,aR,y,z,x)$;
\hspace{1.23cm}
\begin{tikzpicture}[scale = 0.1]
\draw (-1,0) -- (1,0);
\draw (11,0) -- (13,0);
\draw (23,0) -- (25,0);
\draw (35,0) -- (37,0);
\draw (47,0) -- (49,0);
\draw (59,0) -- (61,0);
\V
\begin{scope}[shift={(12,0)}]
\V
\begin{scope}[shift={(-3,4)}, scale=0.5]
\wt
\end{scope}
\begin{scope}[shift={(3,4)}, scale=0.5]
\wt
\end{scope}
\end{scope}

\begin{scope}[shift={(24,0)}]
\V
\begin{scope}[shift={(-3,4)}, scale=0.5]
\wt
\end{scope}
\begin{scope}[shift={(3,4)}, scale=0.5]
\wt
\end{scope}
\end{scope}

\begin{scope}[shift={(48,0)}, scale=0.5]
\wt
\end{scope}

\begin{scope}[shift={(60,0)}, scale=0.5]
\wt
\end{scope}
\end{tikzpicture}\\

\hspace{0.28cm}move$(a,x)$;
\hspace{3.3cm}
\begin{tikzpicture}[scale = 0.1]
\draw (-1,0) -- (1,0);
\draw (11,0) -- (13,0);
\draw (23,0) -- (25,0);
\draw (35,0) -- (37,0);
\draw (47,0) -- (49,0);
\draw (59,0) -- (61,0);
\begin{scope}[shift={(12,0)}]
\V
\begin{scope}[shift={(-3,4)}, scale=0.5]
\wt
\end{scope}
\begin{scope}[shift={(3,4)}, scale=0.5]
\wt
\end{scope}
\end{scope}

\begin{scope}[shift={(24,0)}]
\V
\begin{scope}[shift={(-3,4)}, scale=0.5]
\wt
\end{scope}
\begin{scope}[shift={(3,4)}, scale=0.5]
\wt
\end{scope}
\end{scope}

\begin{scope}[shift={(36,0)}]
\V
\end{scope}

\begin{scope}[shift={(48,0)}, scale=0.5]
\wt
\end{scope}
\begin{scope}[shift={(60,0)}, scale=0.5]
\wt
\end{scope}
\end{tikzpicture}\\

\hspace{0.3cm}$h(n-1,bL,bR,y,a,xL,xR)$;
\hspace{0.65cm}
\begin{tikzpicture}[scale = 0.1]
\draw (-1,0) -- (1,0);
\draw (11,0) -- (13,0);
\draw (23,0) -- (25,0);
\draw (35,0) -- (37,0);
\draw (47,0) -- (49,0);
\draw (59,0) -- (61,0);
\begin{scope}[scale=0.5]
\wt
\end{scope}

\begin{scope}[shift={(12,0)}]
\V
\end{scope}

\begin{scope}[shift={(24,0)}]
\V
\begin{scope}[shift={(-3,4)}, scale=0.5]
\wt
\end{scope}
\begin{scope}[shift={(3,4)}, scale=0.5]
\wt
\end{scope}
\end{scope}

\begin{scope}[shift={(36,0)}]
\V
\begin{scope}[shift={(-3,4)}, scale=0.5]
\wt
\end{scope}
\begin{scope}[shift={(3,4)}, scale=0.5]
\wt
\end{scope}
\end{scope}

\begin{scope}[shift={(60,0)}, scale=0.5]
\wt
\end{scope}
\end{tikzpicture}\\

\hspace{0.3cm}move$(b,y)$;
\hspace{3.3cm}
\begin{tikzpicture}[scale = 0.1]
\draw (-1,0) -- (1,0);
\draw (11,0) -- (13,0);
\draw (23,0) -- (25,0);
\draw (35,0) -- (37,0);
\draw (47,0) -- (49,0);
\draw (59,0) -- (61,0);
\begin{scope}[scale=0.5]
\wt
\end{scope}

\begin{scope}[shift={(24,0)}]
\V
\begin{scope}[shift={(-3,4)}, scale=0.5]
\wt
\end{scope}
\begin{scope}[shift={(3,4)}, scale=0.5]
\wt
\end{scope}
\end{scope}

\begin{scope}[shift={(36,0)}]
\V
\begin{scope}[shift={(-3,4)}, scale=0.5]
\wt
\end{scope}
\begin{scope}[shift={(3,4)}, scale=0.5]
\wt
\end{scope}
\end{scope}

\begin{scope}[shift={(48,0)}]
\V
\end{scope}

\begin{scope}[shift={(60,0)}, scale=0.5]
\wt
\end{scope}
\end{tikzpicture}\\

\hspace{0.3cm}$h(n-1,cL,cR,z,b,yL,yR)$;
\hspace{0.73cm}
\begin{tikzpicture}[scale = 0.1]
\draw (-1,0) -- (1,0);
\draw (11,0) -- (13,0);
\draw (23,0) -- (25,0);
\draw (35,0) -- (37,0);
\draw (47,0) -- (49,0);
\draw (59,0) -- (61,0);
\begin{scope}[scale=0.5]
\wt
\end{scope}

\begin{scope}[shift={(12,0)}, scale=0.5]
\wt
\end{scope}

\begin{scope}[shift={(24,0)}]
\V
\end{scope}

\begin{scope}[shift={(36,0)}]
\V
\begin{scope}[shift={(-3,4)}, scale=0.5]
\wt
\end{scope}
\begin{scope}[shift={(3,4)}, scale=0.5]
\wt
\end{scope}
\end{scope}

\begin{scope}[shift={(48,0)}]
\V
\begin{scope}[shift={(-3,4)}, scale=0.5]
\wt
\end{scope}
\begin{scope}[shift={(3,4)}, scale=0.5]
\wt
\end{scope}
\end{scope}
\end{tikzpicture}\\

\hspace{0.3cm}move$(c,z)$;
\hspace{3.3cm}
\begin{tikzpicture}[scale = 0.1]
\draw (-1,0) -- (1,0);
\draw (11,0) -- (13,0);
\draw (23,0) -- (25,0);
\draw (35,0) -- (37,0);
\draw (47,0) -- (49,0);
\draw (59,0) -- (61,0);
\begin{scope}[scale=0.5]
\wt
\end{scope}

\begin{scope}[shift={(12,0)}, scale=0.5]
\wt
\end{scope}

\begin{scope}[shift={(36,0)}]
\V
\begin{scope}[shift={(-3,4)}, scale=0.5]
\wt
\end{scope}
\begin{scope}[shift={(3,4)}, scale=0.5]
\wt
\end{scope}
\end{scope}

\begin{scope}[shift={(48,0)}]
\V
\begin{scope}[shift={(-3,4)}, scale=0.5]
\wt
\end{scope}
\begin{scope}[shift={(3,4)}, scale=0.5]
\wt
\end{scope}
\end{scope}

\begin{scope}[shift={(60,0)}]
\V
\end{scope}
\end{tikzpicture}\\

\hspace{0.3cm}$g(n-1,a,b,zL,zR,c)$;
\hspace{1.65cm}
\begin{tikzpicture}[scale = 0.1]
\draw (-1,0) -- (1,0);
\draw (11,0) -- (13,0);
\draw (23,0) -- (25,0);
\draw (35,0) -- (37,0);
\draw (47,0) -- (49,0);
\draw (59,0) -- (61,0);
\begin{scope}[shift={(36,0)}]
\V
\begin{scope}[shift={(-3,4)}, scale=0.5]
\wt
\end{scope}
\begin{scope}[shift={(3,4)}, scale=0.5]
\wt
\end{scope}
\end{scope}

\begin{scope}[shift={(48,0)}]
\V
\begin{scope}[shift={(-3,4)}, scale=0.5]
\wt
\end{scope}
\begin{scope}[shift={(3,4)}, scale=0.5]
\wt
\end{scope}
\end{scope}

\begin{scope}[shift={(60,0)}]
\V
\begin{scope}[shift={(-3,4)}, scale=0.5]
\wt
\end{scope}
\begin{scope}[shift={(3,4)}, scale=0.5]
\wt
\end{scope}
\end{scope}
\end{tikzpicture}\\

{\bf end}

\smallskip
\noindent
{\bf end.}

\bigskip
Let $f_n$, $g_n$, and $h_n$ denote the number of moves executed by 
calls of $f$, $g$, and $h$, respectively, with trees of height $n$. 
Clearly $f_0=g_0=h_0=0$ and, for $n\geq 1$, 
\begin{enumerate}
\item[(1)] $f_n=2g_{n-1}+1$
\item[(2)] $g_n= 2g_{n-1}+h_{n-1}+2$
\item[(3)] $h_n= 2g_{n-1}+2h_{n-1}+3$. 
\end{enumerate}
Solving $h_{n-1}$ in (2) and substituting it in (3) gives 
%
$$h_n=2g_{n-1}+2(g_n-2g_{n-1}-2)+3=2g_n-2g_{n-1}-1,$$
and so $h_{n-1}=2g_{n-1}-2g_{n-2}-1$. Substituting this back into~(2) gives 
%
$$g_n=2g_{n-1}+(2g_{n-1}-2g_{n-2}-1)+2=4g_{n-1}-2g_{n-2}+1$$
and so $g_{n-1}=4g_{n-2}-2g_{n-3}+1$.
Hence 
%
$$2g_{n-1}+1=4(2g_{n-2}+1)-2(2g_{n-3}+1)+1$$
and so by (1)
\begin{enumerate}
\item[(4)] $f_n=4f_{n-1}-2f_{n-2}+1$ $\quad$ (for $n\geq 2$; \;$f_0=0$, $f_1=1$).
\end{enumerate}
To solve (4) consider $p_n=f_n+1$. Then 
\begin{enumerate}
\item[(5)] $p_0=1$, $p_1=2$, and $p_n=4p_{n-1}-2p_{n-2}$ for $n\geq 2$.
\end{enumerate}
To solve the (linear homogeneous) recurrence relation (5), 
let $G(z) =\sum_{i=0}^\infty \,p_iz^i$ be the generating function of the sequence $p_i$. 
From (5) we obtain 
$$(1-4z+2z^2)\,G(z)=1-2z.$$ 
Hence 
$$G(z)= \frac{1-2z}{1-4z+2z^2}=\frac{1}{2}(\frac{1}{1-\tau z}+\frac{1}{1-\hat{\tau} z})$$
where $\tau=2+\sqrt{2}$ and $\hat{\tau}=2-\sqrt{2}$. 
Consequently, 
$$G(z)= \frac{1}{2}(1+\tau z+\tau^2z^2+ \cdots +1+\hat{\tau} z+\hat{\tau}^2z^2+ \cdots)$$
and so $p_n=\frac{1}{2}(\tau^n+\hat{\tau}^n)$. 
Thus, for $n\geq 0$, 
\begin{enumerate}
\item[(6)] $f_n= \frac{1}{2}(2+\sqrt{2})^n + \frac{1}{2}(2-\sqrt{2})^n -1$. 
\end{enumerate}
Hence $f_n = \lfloor\frac{1}{2}(2+\sqrt{2})^n\rfloor$ because 
$(2-\sqrt{2})^n$ is small ($2-\sqrt{2}\approx 0.58$).
Note that $2+\sqrt{2}\approx 3.42$ and so $f_n$ is an improvement on $t_n$. 

\smallskip
In the remainder of the paper we will show that (6) is optimal: 
any sequence of moves leading from the initial configuration to the final one 
(with the full binary tree of height $n$) contains at least $f_n$ moves. 
The proof goes by showing simultaneously the analogous statements for $g_n$ and $h_n$ 
by induction on $n$, using the formulas (1)\,--\,(3).  For $n=0$ (or $n=1$) this is obvious. 
Assume we have shown optimality of $f_{n-1}$, $g_{n-1}$ and $h_{n-1}$, and let us prove it for $n$. 
For each case ($f_n$, $g_n$, $h_n$) we consider a sequence of moves from the initial to the final 
configuration with trees of height $n$.

First the ``$f$-case'', i.e., one tree and four places. This uses exactly the same argument as in 
the ordinary Hanoi-case (cf. the case of $t$). Consider the first move of the largest node; 
at that moment the two subtrees of height $n-1$ have been moved to the other two places 
(without moving the largest node). Hence, by induction, this took at least $g_{n-1}$ moves. 
By similarly considering the last move of the largest node, it follows that the total number of moves 
is at least $g_{n-1}+1+g_{n-1}=2g_{n-1}+1=f_n$.

\begin{figure}[t]
\begin{tikzpicture}[scale = 0.14]
\draw (-1,0) -- (1,0);
\draw (11,0) -- (13,0);
\draw (35,0) -- (37,0);
\draw (47,0) -- (49,0);
\draw (59,0) -- (61,0);
\draw (71,0) -- (73,0);
\V
\begin{scope}[shift={(-3,4)}, scale=0.5]
\wt
\end{scope}
\begin{scope}[shift={(3,4)}, scale=0.5]
\wt
\end{scope}
\begin{scope}[shift={(12,0)}]
\V
\begin{scope}[shift={(-3,4)}, scale=0.5]
\wt
\end{scope}
\begin{scope}[shift={(3,4)}, scale=0.5]
\wt
\end{scope}
\end{scope}

\node [above] at (24,1) {$\cdots\cdot$};

\begin{scope}[shift={(36,0)}]
\V
\begin{scope}[shift={(-3,4)}, scale=0.5]
\wt
\end{scope}
\begin{scope}[shift={(3,4)}, scale=0.5]
\wt
\end{scope}
\end{scope}

\node [below] at (18,-3) {$\underbrace{\hspace{5.5cm}}$};
\node [below] at (18,-7) {``old'' places};

\node [below] at (60,-3) {$\underbrace{\hspace{3.8cm}}$};
\node [below] at (60,-7) {``new'' places};

\end{tikzpicture}
  \caption{General initial configuration.}
  \label{fig:genconf}
\end{figure}
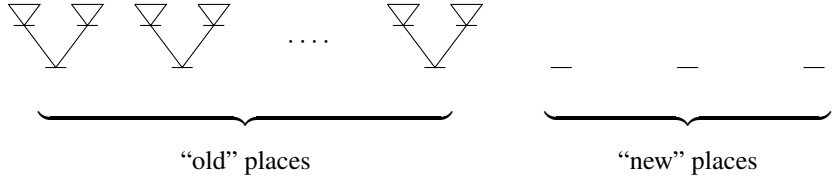

Before proving the ``$g$-case'' and ``$h$-case'' we need a few general observations. 
Consider a more general initial configuration as in Fig.~\ref{fig:genconf}, 
where we have any number of full binary trees of height $n$, 
of which $k$ (with $k=2$ or $3$) have to be moved to $k$ of the three empty places.
Let us say that two configurations $c_1$ and $c_2$ of this game are \emph{equivalent}
if $c_1$ can be obtained from $c_2$ by permuting the old places among each other 
and also permuting the new places among each other (see Fig.~\ref{fig:genconf}
for the meaning of ``old'' and ``new''). 
Clearly, if $c_1$ and $c_2$ are equivalent and $c_1$ can be reached 
from the initial configuration $c_0$ in $m$ moves, then $c_2$ can also be reached 
from $c_0$ in $m$ moves (permute the place numbers in the positions in the moves). 
Hence, in a shortest sequence of moves from $c_0$ to the final configuration $c_\infty$,
there do not occur two equivalent configurations (if 
$c_0 \stackrel{m}{\Rightarrow} c_1 \stackrel{s}{\Rightarrow} c_2 \stackrel{t}{\Rightarrow} c_\infty$
and $c_1,c_2$ are equivalent, then $c_0 \stackrel{m}{\Rightarrow} c_1 \stackrel{t}{\Rightarrow} c_\infty$). 
Now consider, in such a shortest sequence, the moves of the largest nodes. 
At the moment that a largest node moves from one place to another, 
two of the other places are filled with a full tree of height $n-1$
and the remaining places are filled with full trees of height $n$.
The first remark is that a large node never moves 
from an old place to an old place, or from a new place to a new place. 
In fact, the configurations just before and after such a move are equivalent. The second remark is 
that if a large node moves from an old to a new place, then the next large node 
will \emph{not} move from a new to an old place. Suppose it does. 
Just after the first large node moved from an old place to a new one, the configuration is, e.g., 

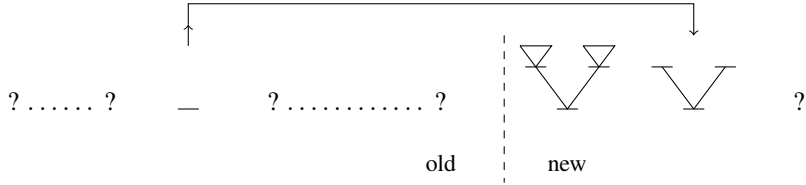
\bigskip
\begin{tikzpicture}[scale = 0.14]

\draw [->] (12,6) -- (12,8);
\draw [->] (12,8) -- (12,10) -- (60,10) -- (60,7);


\node [above] at (0,-1) {? $\dots\dots$ ?};

\draw (11,0) -- (13,0);


\node [above] at (28,-1) {? $\dots\dots\dots\dots$ ?};


\draw [dashed] (42,7) -- (42,-7);

\draw (47,0) -- (49,0);
\draw (59,0) -- (61,0);

\node[above] at (70,-1) {?};



\begin{scope}[shift={(48,0)}]
\V
\begin{scope}[shift={(-3,4)}, scale=0.5]
\wt
\end{scope}
\begin{scope}[shift={(3,4)}, scale=0.5]
\wt
\end{scope}
\end{scope}

\begin{scope}[shift={(60,0)}]
\V
\end{scope}

\node [below] at (36,-3.5) {{\small old}};
\node [below] at (48,-4) {{\small new}};

\end{tikzpicture}

\medskip
\noindent
where each ? is either a full tree of height $n-1$ or a full tree of height $n$. 
Just before the next large node moves, the configuration is, e.g., 

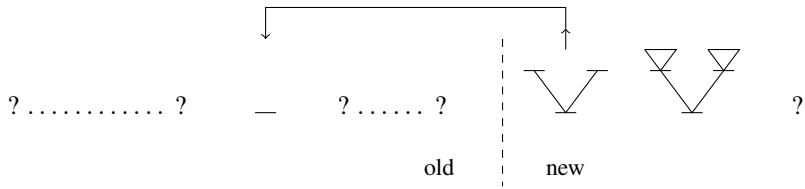
\bigskip
\begin{tikzpicture}[scale = 0.14]

\draw [->] (52.5,6) -- (52.5,8);
\draw [->] (52.5,8) -- (52.5,10) -- (24,10) -- (24,7);


\node [above] at (8,-1) {? $\dots\dots\dots\dots$ ?};

\draw (23,0) -- (25,0);


\node [above] at (36,-1) {? $\dots\dots$ ?};


\draw [dashed] (46.5,7) -- (46.5,-7);

\draw (51.5,0) -- (53.5,0);
\draw (63.5,0) -- (65.5,0);

\node[above] at (74.5,-1) {?};



\begin{scope}[shift={(52.5,0)}]
\V
\end{scope}

\begin{scope}[shift={(64.5,0)}]
\V
\begin{scope}[shift={(-3,4)}, scale=0.5]
\wt
\end{scope}
\begin{scope}[shift={(3,4)}, scale=0.5]
\wt
\end{scope}
\end{scope}

\node [below] at (40.5,-3.5) {{\small old}};
\node [below] at (52.5,-4) {{\small new}};

\end{tikzpicture}

\medskip
\noindent
where the empty (old) place may be different. Since inbetween these configurations 
no large node has been moved, these two configurations are equivalent, which is a contradiction
(note that the two configurations are not the same, because in that case the same large node 
immediately moves back to the same place, which does not happen in a shortest sequence of moves). 
We can conclude that, in a shortest sequence of moves, the largest nodes always move 
from an old place to a new place, and hence 
only $k$ such moves ($k=2$ or $3$) occur in the sequence. 
 
We now show the ``$g$-case'', i.e., $k=2$. Consider, for a shortest sequence of moves, 
the initial configuration, the configurations at the time of the two moves of the largest nodes, 
and the final configuration, as follows. 

\bigskip
$c_0$
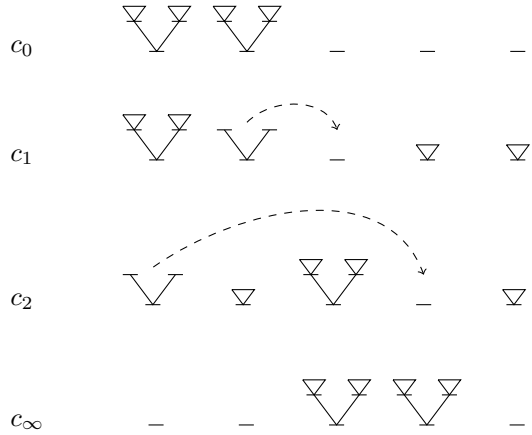
\hspace{1cm}
\begin{tikzpicture}[scale = 0.1]
\draw (-1,0) -- (1,0);
\draw (11,0) -- (13,0);
\draw (23,0) -- (25,0);
\draw (35,0) -- (37,0);
\draw (47,0) -- (49,0);
\V
\begin{scope}[shift={(-3,4)}, scale=0.5]
\wt
\end{scope}
\begin{scope}[shift={(3,4)}, scale=0.5]
\wt
\end{scope}
\begin{scope}[shift={(12,0)}]
\V
\begin{scope}[shift={(-3,4)}, scale=0.5]
\wt
\end{scope}
\begin{scope}[shift={(3,4)}, scale=0.5]
\wt
\end{scope}
\end{scope}
\end{tikzpicture}

\vspace{0.5cm}
$c_1$
\hspace{1cm}
\begin{tikzpicture}[scale = 0.1]
\draw (-1,0) -- (1,0);
\draw (11,0) -- (13,0);
\draw (23,0) -- (25,0);
\draw (35,0) -- (37,0);
\draw (47,0) -- (49,0);
\V
\begin{scope}[shift={(-3,4)}, scale=0.5]
\wt
\end{scope}
\begin{scope}[shift={(3,4)}, scale=0.5]
\wt
\end{scope}
\begin{scope}[shift={(12,0)}]
\V
\end{scope}

\begin{scope}[shift={(36,0)}, scale=0.5]
\wt
\end{scope}

\begin{scope}[shift={(48,0)}, scale=0.5]
\wt
\end{scope}

\draw[dashed,->] (12,5) to [out=45, in=110] (24,4);
\end{tikzpicture}

\vspace{0.1cm}
$c_2$
\hspace{1cm}
\begin{tikzpicture}[scale = 0.1]
\draw (-1,0) -- (1,0);
\draw (11,0) -- (13,0);
\draw (23,0) -- (25,0);
\draw (35,0) -- (37,0);
\draw (47,0) -- (49,0);
\V

\begin{scope}[shift={(12,0)}, scale=0.5]
\wt
\end{scope}

\begin{scope}[shift={(24,0)}]
\V
\begin{scope}[shift={(-3,4)}, scale=0.5]
\wt
\end{scope}
\begin{scope}[shift={(3,4)}, scale=0.5]
\wt
\end{scope}
\end{scope}

\begin{scope}[shift={(48,0)}, scale=0.5]
\wt
\end{scope}

\draw[dashed,->] (0.05,5) to [out=35, in=110] (36,4);
\end{tikzpicture}

\vspace{0.9cm}
$c_\infty$
\hspace{1.2cm}
\begin{tikzpicture}[scale = 0.1]
\draw (-1,0) -- (1,0);
\draw (11,0) -- (13,0);
\draw (23,0) -- (25,0);
\draw (35,0) -- (37,0);
\draw (47,0) -- (49,0);

\begin{scope}[shift={(24,0)}]
\V
\begin{scope}[shift={(-3,4)}, scale=0.5]
\wt
\end{scope}
\begin{scope}[shift={(3,4)}, scale=0.5]
\wt
\end{scope}
\end{scope}

\begin{scope}[shift={(36,0)}]
\V
\begin{scope}[shift={(-3,4)}, scale=0.5]
\wt
\end{scope}
\begin{scope}[shift={(3,4)}, scale=0.5]
\wt
\end{scope}
\end{scope}
\end{tikzpicture}

\medskip
\noindent
From $c_0$ to $c_1$, two trees of height $n-1$ are moved to new positions. Hence, by induction,
at least $g_{n-1}$ moves are made. Similarly, from $c_1$'s successor to $c_2$, three trees of height $n-1$
are moved, which (by induction) takes at least $h_{n-1}$ moves. Finally, from $c_2$'s successor to $c_\infty$,
two trees of height $n-1$ are moved, taking at least $g_{n-1}$ moves. Hence the number of moves in 
such a shortest sequence is at least $g_{n-1}+1+h_{n-1}+1+g_{n-1}=2g_{n-1}+h_{n-1}+2=g_n$. 
Note that, formally, we actually have to consider the more general case of Fig.~\ref{fig:genconf}: 
from $c_0$ to $c_1$ two trees of height $n-1$ are moved, 
but the other two subtrees of height $n-1$ are also movable! 
Hence we use the induction hypothesis for the case of Fig.~\ref{fig:genconf} 
with four old places (and $k=2$). It should be clear that the proof of this more general case 
goes in exactly the same way as above. 

The ``$h$-case'', i.e., $k=3$. In the whole (shortest) sequence of moves, each of the three largest nodes 
moves exactly once (from an old to a new place). By considering the configurations at these moments 
(together with the initial and final configurations) it is easy to see that the number of moves is at least 
$g_{n-1}+1+h_{n-1}+1+h_{n-1}+1+g_{n-1}=2g_{n-1}+2h_{n-1}+3=h_n$.

This finally shows the optimality of formulas (1) to (3), and hence of formula (6). 
Thus the procedure $f$ uses the minimal number of moves to solve our generalized Hanoi problem. 

We mention the following two questions.

\begin{enumerate}
\item[(1)] Does there exist an easy iterative (non-recursive) solution of the Trees of Hanoi? 
cf.~\cite{hay77} for the Towers of Hanoi. 
\item [(2)] How should the game be played in the case of an arbitrary binary tree 
in the initial configuration?
\end{enumerate}

\bigskip
\noindent
{\bf Acknowledgement.}
I thank Maarten Fokkinga for his help.

\vspace{1.5cm}
\begin{center}
\begin{tikzpicture}
\begin{scope}[scale=0.8]
\draw (0,2) -- (1.5,0) -- (3,2);
\draw (1,0) -- (2,0);
\draw (-0.5,2) -- (0.5,2);
\draw (2.5,2) -- (3.5,2);
\end{scope}
\end{tikzpicture}
\end{center}

\newpage
\section*{Epilogue}

The previous text is a slightly revised version 
of a note that I wrote 36 years ago~\cite{eng81}. 
A few errors were corrected and a few things improved. 

In~\cite{jurwoo83} the game of the Trees of Hanoi was generalized to full $m$-ary trees
instead of full binary trees, for any $m\geq 1$, with $m+2$ places.  
In fact, the generalization is a straightforward adaptation of the above case $m=2$.  
The minimal solution still consists of three recursive procedures $f$, $g$, and $h$, but now
$g$ moves $m$ trees and $h$ moves $m+1$ trees from old to new places. 
Thus, $f_n=2g_{n-1}+1$ as before, but now 
$g_n=2g_{n-1}+(m-1)h_{n-1}+m$ and $h_n=2g_{n-1}+mh_{n-1}+m+1$. 
This leads to $f_n= (m+2)f_{n-1} -2f_{n-2}+m-1$ and to 
$f_n= \lfloor \frac{1}{2R}(R-m+2)\tau^n\rfloor$ 
where $R=\sqrt{(m+2)^2-8}$ and $\tau=\frac{1}{2}(m+2+R)$. 
In the optimality proof, 
the general initial configuration of Fig.~\ref{fig:genconf} should have $m+1$ empty new places
and any number of full $m$-ary trees of height $n$ on the old places, of which $k$ 
have to be moved to the new places, with $k=m$ or $m+1$.

When I left the Theoretical Computer Science group of Twente University of Technology in 1984, 
I received a physical model of the Trees of Hanoi, nicely constructed by Maarten and Ans Fokkinga. 
With it, I could physically demonstrate the game to my students, colleagues, friends, and family,
not only the (binary) Trees of Hanoi, but also the ternary Trees of Hanoi and the classical Towers of Hanoi.
The model consists of silver-coloured cylinders of five different sizes, such that 
on each cylinder of a given size at most three cylinders can be placed of the next smaller size. 
To allow quick moves, the smaller cylinders are not attached to the larger one in any way, 
but just stand freely on top of it. 
Using all cylinders, a full ternary tree of height~5 can be built, with 81 very small cylinders as leaves.  
This tree can be moved from one place to another in 323 moves,
or in one big move if your hands are not shaking.

\end{document}